\documentclass[12pt]{article}
\usepackage{amssymb}

\def\be{\begin{equation}}
\def\ee{\end{equation}}
\def\bea{\begin{eqnarray}}
\def\eea{\end{eqnarray}}
\def\ba{\begin{array}}
\def\ea{\end{array}}
\def\bi{\begin{itemize}}
\def\ei{\end{itemize}}
\def\half{{\textstyle{1\over2}}}

\catcode`\@=11
\def\@citex[#1]#2{%
\if@filesw \immediate \write \@auxout {\string \citation {#2}}\fi
\@tempcntb\m@ne \let\@h@ld\relax \def\@citea{}%
\@cite{%
  \@for \@citeb:=#2\do {%
    \@ifundefined {b@\@citeb}%
      {\@h@ld\@citea\@tempcntb\m@ne{\bf ?}%
      \@warning {Citation `\@citeb ' on page \thepage \space undefined}}%
      {\@tempcnta\@tempcntb \advance\@tempcnta\@ne%
      \@tempcntb\number\csname b@\@citeb \endcsname \relax%
      \ifnum\@tempcnta=\@tempcntb 
        \ifx\@h@ld\relax%
          \edef \@h@ld{\@citea\csname b@\@citeb\endcsname}%
        \else%
          \edef\@h@ld{\ifmmode{-}\else--\fi\csname b@\@citeb\endcsname}%
        \fi%
      \else
        \@h@ld\@citea\csname b@\@citeb \endcsname%
        \let\@h@ld\relax%
      \fi}%
    \def\@citea{,\penalty\@highpenalty\,}%
  }\@h@ld
}{#1}}

\def\@citeb#1#2{{[#1]\if@tempswa , #2\fi}}
\def\@citeu#1#2{{$^{#1}$\if@tempswa , #2\fi }}
\def\@citep#1#2{{#1\if@tempswa , #2\fi}}

\def\bcites{         
        \catcode`\@=11
        \let\@cite=\@citeb
        \catcode`\@=12
}

\def\upcites{         
        \catcode`\@=11
        \let\@cite=\@citeu
        \catcode`\@=12
}

\def\plaincites{      
        \catcode`\@=11
        \let\@cite=\@citep
        \catcode`\@=12
}

\newcount\hour
\newcount\minute
\newtoks\amorpm
\hour=\time\divide\hour by 60
\minute=\time{\multiply\hour by 60 \global\advance\minute by-\hour}
\edef\standardtime{{\ifnum\hour<12 \global\amorpm={am}%
        \else\global\amorpm={pm}\advance\hour by-12 \fi
        \ifnum\hour=0 \hour=12 \fi
        \number\hour:\ifnum\minute<10 0\fi\number\minute\the\amorpm}}
\edef\militarytime{\number\hour:\ifnum\minute<10 0\fi\number\minute}

\def\draftlabel#1{{\@bsphack\if@filesw {\let\thepage\relax
   \xdef\@gtempa{\write\@auxout{\string
      \newlabel{#1}{{\@currentlabel}{\thepage}}}}}\@gtempa
   \if@nobreak \ifvmode\nobreak\fi\fi\fi\@esphack}
        \gdef\@eqnlabel{#1}}
\def\@eqnlabel{}
\def\@vacuum{}
\def\marginnote#1{}
\def\draftmarginnote#1{\marginpar{\raggedright\scriptsize\tt#1}}
\def\draft{
        \pagestyle{plain}
        \overfullrule=2pt
        \oddsidemargin -.5truein
        \def\@oddhead{\sl \phantom{\today\quad\militarytime} \hfil
        \smash{\Large\sl DRAFT} \hfil \today\quad\militarytime}
        \let\@evenhead\@oddhead
        \let\label=\draftlabel
        \let\marginnote=\draftmarginnote
        \def\ps@empty{\let\@mkboth\@gobbletwo
        \def\@oddfoot{\hfil \smash{\Large\sl DRAFT} \hfil}
        \let\@evenfoot\@oddhead}
        \def\@eqnnum{(\theequation)\rlap{\kern\marginparsep\tt\@eqnlabel}%
        \global\let\@eqnlabel\@vacuum}  }

\begin{document}


\hfill UTHET-02-0101

\vspace{-0.2cm}

\begin{center}
\Large
{ \bf An AdS/dS duality for a scalar particle\footnote{Research supported by the DoE under grant DE-FG05-91ER40627.}}
\normalsize

\vspace{0.8cm}
 
{\bf George Siopsis}\footnote{gsiopsis@utk.edu}\\ Department of Physics
and Astronomy, \\
The University of Tennessee, Knoxville, \\
TN 37996 - 1200, USA.
 \end{center}

\vspace{0.8cm}
\large
\centerline{\bf Abstract}
\normalsize
\vspace{.5cm}

The motion of a scalar particle in $(d+1)$-dimensional AdS space may be
described
in terms of the Cartesian coordinates that span the $(d+2)$-dimensional space
in which the AdS space is embedded. Upon quantization, the mass hyperboloid defined in terms
of the conjugate momenta turns into the wave equation in AdS space.
By interchanging the roles of coordinates and conjugate momenta in the
$(d+2)$-dimensional space we arrive at a dual description. For
massive modes, the dual description is equivalent to the conventional formulation, as required by holography. For tachyonic modes,
this interchange of coordinates and momenta establishes a duality between
Euclidean AdS and dS spaces. We discuss its implications on Green functions for the various
vacua.

\newpage

Although a lot of similarities exist between AdS and dS spaces, they differ
in significant ways. Perhaps most importantly, dS spaces have not been
understood in a
string-theoretical framework, making it impossible to obtain their thermodynamic properties, such as entropy, microscopically.
A significant step in this direction was the recent proposal by Strominger~\cite{strominger}
of a dS/CFT correspondence where the CFT lies in the infinite past of dS space.
Naturally, it attracted much attention~\cite{bibds1,bibds2,bibds3,bibds4,bibds5,bibds6,bibds7,bibds8,bibds9,bibds10,bibds11,bibds12,bibds13,bibds14,bibds15,bibds16,bibds17,bibds18,bibds19,bibds20,bibds21,bibds22,bibds23,bibds24,bibds25,bibds26,bibds27,bibds28,bibds29,bibds30}.

The dS/CFT correspondence bears a striking formal resemblance to its AdS counterpart~\cite{bibads1,bibads2,bibads3} suggesting that the former be derived from the latter by some kind of
analytic continuation~\cite{bibadsds,bibadsds2,bibadsds3}. As was pointed out in~\cite{bibbms}, one needs to exercise care in
such extrapolations. If one analyzes the behavior of the respective Green
functions carefully, one discovers that dS Green functions may not be obtained
by a double analytic continuation of their AdS counterparts.

Here we discuss a different proposal of extrapolating from AdS to dS spaces.
We establish a duality between the two spaces which interchanges the role
of coordinates and momenta for a scalar field. We thus show that a massive
mode in dS space is dual to a tachyonic mode in AdS space. This is based
on the following basic observation. A $(d+1)$-dimensional AdS space (AdS$_{d+1}$)
is defined within a flat $(d+2)$-dimensional space as the hypersurface
\be\label{eqc1}
X_0^2 - X_1^2 - \dots - X_d^2 + X_{d+1}^2 = \ell^2
\ee
where $\ell^2 > 0$.
A particle of mass $m$ moving in this space has a trajectory on the mass-shell
hypersurface in momentum space
\be\label{eqc2}
P_0^2 - P_1^2  - \dots - P_d^2 + P_{d+1}^2 = m^2
\ee
After an analytic continuation, $X_{d+1}\to iX_{d+1}$ (and correspondingly,
$P_{d+1}\to -iP_{d+1}$), one obtains the Euclidean AdS space (EAdS).
The mass-shell condition then reads
\be\label{eqc2ds}
-P_0^2 + P_1^2  + \dots + P_d^2 + P_{d+1}^2 = -m^2
\ee
The form of this constraint~(\ref{eqc2ds}) is identical to the constraint~(\ref{eqc1}) for massive modes ($m^2> 0$). Thus, if we interchange the role
of coordinates and momenta, we arrive at the same theory in (E)AdS space.

On the other hand, for tachyonic modes ($m^2<0$)~\cite{bibtach1,bibtach2}, eq.~(\ref{eqc2ds})
is the defining equation of the dS hyperboloid. Thus, by interchanging the
roles of momenta and coordinates in the $(d+2)$-dimensional space, we establish
a duality between tachyonic modes in EAdS space and massive scalars in dS space.
This is our main result, which we now proceed to discuss in some detail.

Let us start with the embedding $(d+2)$-dimensional space with flat metric given by
\be\label{eqmet3}
ds^2 = dX_AdX^A = - dX_0^2 + dX_1^2 + \dots + dX_d^2 - dX_{d+1}^2
\ee
and introduce the conjugate momenta $P_A$ ($A=0,1,\dots,d+1$). A particle
of mass $m$ moving in the embedding will obey the mass-shell condition~(\ref{eqc2ds}). Upon restricting its motion to
AdS$_{d+1}$ space, which is the hypersurface~(\ref{eqc1}), the mass-shell
condition is given in terms of the Casimir
\be\label{eqcas} \mathcal{C}_2 = \half J_{AB}J^{AB}\ee
where we introduced the angular momentum
\be J_{AB} = X_AP_B - X_BP_A\ee 
To impose the restriction onto AdS$_{d+1}$ space, we introduce the constraint
\be\label{eqcon3} \mathcal{D} \equiv X^AP_A = 0\ee
where $\mathcal{D}$ is the generator of scale transformations in the embedding.
Notice that the Casimir~(\ref{eqcas}) is invariant under scale transformations
and therefore its Poisson bracket with $\mathcal{D}$ vanishes. Upon imposing
the constraint~(\ref{eqcon3}), the Casimir may be written as
\be\label{eqcon4} \mathcal{C}_2 = (X_AX^A)(P_BP^B)\ee
The theory defined by~(\ref{eqcas}) and (\ref{eqcon3}) is a gauge theory and
$\mathcal{D}$ is the generator of gauge transformations. To quantize the
system, we need to fix the gauge. We shall do so by imposing the gauge-fixing
condition
\be\label{eqgf3} X_AX^A = -\ell^2\quad, \quad \ell^2 > 0\ee
which restricts motion onto the AdS$_{d+1}$ space with parameter $\ell$. It
should be emphasized that the choice of $\ell$ is arbitrary. Different choices
are related by a gauge (scale) transformation and the theory is solely defined
in terms of the value of the Casimir $\mathcal{C}_2$. The latter may be written
as
\be\label{eqcon5} \mathcal{C}_2 = m^2\ell^2\ee
where $m$ has the standard interpretation of mass ($P_AP^A = -m^2$) on account of~(\ref{eqcon4}).

Next, we compute the Dirac brackets which will then be promoted to commutators.
A simple calculation yields
\be\label{eqdir3} \{ X^A\;,\; P_B\}_D = \delta_B^A + \frac{1}{\ell^2} X^AX_B\ee
This algebra is realized in terms of Poincar\'e coordinates
\bea\label{eqpc}
X_0 & = & \frac z 2 + \frac{x^\mu x_\mu + \ell^2}{2z} \nonumber \\
X^i & = & \frac\ell z\, x^i\quad (i=1,\dots,d-1), \quad X_{d+1} = \frac\ell z\, x_0\nonumber \\
X^d & = & \frac z 2 + \frac{x^\mu x_\mu - \ell^2}{2z}
\eea
where $x^\mu x_\mu = - (x^0)^2 +(x^1)^2 + \dots + (x^{d-1})^2$, which cover half of AdS. The metric~(\ref{eqmet3}) restricted to the AdS hypersurface reads
\be\label{eqmetr}
ds_\mathrm{AdS}^2 = \ell^2\; \frac{dz^2 +dx^\mu dx_\mu}{z^2}
\ee
For future reference, we also express the invariant distance between
points $X^A$ and ${X'}^A$ on the AdS hypersurface in terms of Poincar\'e coordinates:
\be\label{eqP} P(X,X') = \frac{1}{\ell^2}\; (X'-X)^2 =
\frac{(z-z')^2+(x-x')^\mu (x-x')_\mu}{zz'}\ee
The Casimir~(\ref{eqcas}) may also be expressed in terms of the Poincar\'e coordinates
$(z,x^\mu)$~(\ref{eqpc}) and the conjugate momenta $(p_z,p_\mu)$.
Upon quantization, it turns into the 
 Schr\"odinger (wave)
equation
\be\label{eqwe}
z^{d-1}\frac\partial{\partial z} \left( z^{-d+1} \frac{\partial\Psi}{\partial z}\right) + \partial_\mu\partial^\mu\, \Psi = \frac{m^2\ell^2}{z^2}\, \Psi
\ee
where we expressed the value of the Casimir as in~(\ref{eqcon5}).
The inner product in the space of solutions is given by
\be\label{eqip} (\Psi_1,\Psi_2) = i\pi\int_{\Sigma_0} \frac{d^{d-1}x\,dz}{z^{d-1}} (\Psi_1^\star \partial_0
\Psi_2 - \partial_0 \Psi_1^\star \Psi_2) \ee
where $\Sigma_0$ is the spacelike slice $x^0=$~const..
Assuming the wavefunction is a plane wave in the space spanned by $x^\mu$,
\be\label{eqpw} \Psi_k (z,x^\mu) = e^{ik_\mu x^\mu} \Phi_q (z)\ee
where $q^2= -k_\mu k^\mu$, we obtain
\be\label{eqwav}
\Phi_q'' - \frac{d-1} z\; \Phi_q' + q^2\, \Phi_q = \frac{m^2\ell^2}{z^2}\, \Phi_q
\ee
The solution to this equation is written in terms of Bessel functions
\be\label{eqphi}
\Phi_q^\pm (z) = z^{d/2}\; J_{\pm\nu} (qz)\quad,\quad \nu = \sqrt{\frac{d^2}{4} +m^2\ell^2}\ee
The inner product of two wavefunctions is
\be\label{eqnorm} (\Psi_k^\pm,\Psi_{k'}^\pm) = (2\pi)^d \delta^{d-1}(\vec k - \vec k')
\; \frac{k_0+k_0'}{2}\;
\int_0^\infty \frac{dz}{z^{d-1}}\; \Phi_q^{\pm\star} (z)\Phi_{q'}^\pm (z)\ee
where $\vec k = (k_1,\dots, k_{d-1})$ and similarly for $\vec k'$.
At the boundary ($z\to 0$), the two solutions behave as
\be \Phi_q^\pm \sim z^{h_\pm} \quad,\quad h_\pm = \frac{d}{2} \pm \nu\ee
For $m^2>0$, the solution $\Phi_q^-$ is not normalizable, so it is discarded.
The normalizable modes form an orthonormal set with respect to the inner product~(\ref{eqip}),
\be (\Psi_k^+,\Psi_{k'}^+) = (2\pi)^d\, \delta^d(k -k')
\ee
where we used the orthogonality property of Bessel functions,
\be \int_0^\infty dz\, z\, J_\nu (qz) \, J_\nu (q'z) = \frac{1}{q}\, \delta (q-q')\ee
Next, we introduce the propagator
\be\label{eqgpsi} G (z,x^\mu ;z',{x'}^\mu) = \int \frac{d^d k}{(2\pi)^d}\, \Psi_k^{+\star} (z,x^\mu)\Psi_k^+ (z',{x'}^\mu)\ee
which obeys the wave equation~(\ref{eqwe}).
\footnote{We replaced the integration variable $q$ with $k_0$ in order to
arrive at a more convenient expression for the measure in~(\ref{eqgpsi}).}
After some algebra involving Bessel and Hypergeometric function identities,
we arrive at
\be\label{eqG}
G(z,x^\mu; z', {x'}^\mu) = \frac{\Gamma (h_+)}
{2\pi^{d/2}\Gamma (\nu+1)} \; P^{-h_+}\, F(h_+, \nu+\half; 2\nu+1;-4/P)\ee
where the invariant distance $P$ is given by~(\ref{eqP}).
The singularity is obtained by letting $P\to 0$,
\be\label{eqsing} G(z,x^\mu; z', {x'}^\mu) \sim \frac{\Gamma((d-1)/2)}
{4\pi^{(d+1)/2}} \; P_\epsilon^{-(d-1)/2}\ee
where $P_\epsilon$ includes the $i\epsilon$ prescription $x^0-{x'}^0 \to
x^0-{x'}^0 -i\epsilon$.
The Feynman propagator is
\be G_F (z,x^\mu;z',{x'}^\mu) =\theta(x^0-{x'}^0) G (z,x^\mu;z',{x'}^\mu) +
\theta({x'}^0-x^0) G (z',{x'}^\mu;z,x^\mu)\ee
For completeness, we also derive
the bulk-to-boundary propagator, which is obtained by letting one of the arguments
approach the boundary.
In the limit $z'\to 0$, we have $G(z,x^\mu; z', {x'}^\mu) \sim P^{-h_+}$, so
\be G(z,x^\mu; z' {x'}^\mu) \to \frac{1}{2\nu}\; {z'}^{h_+}\; K(z,x^\mu;{x'}^\mu)\ee
where
\be\label{eqgf} K(z,x^\mu;{x'}^\mu) = \pi^{-d/2}\; \frac{\Gamma(h_+)}{\Gamma (\nu)}\;
\left( \frac{z}{z^2+(x-x')^\mu (x-x')_\mu}\right)^{h_+}\ee
In the limit $z\to 0$, this leads to a propagator of the form
\be \Delta (x) \sim (x^\mu x_\mu)^{-h_+}\ee
which is the two-point function (up to a constant) of the corresponding conformal field theory.


The above construction for massive modes carries over to the $m^2 < 0$ regime
(negative Casimir, $\mathcal{C}_2 < 0$; see~(\ref{eqcon5})).
In this regime,
both solutions $\Phi_q^\pm$ (eq.~(\ref{eqphi})) may be acceptable leading to
distinct theories and therefore different Green functions hinting at symmetry breaking. Boundary conditions
select one of the possible propagators. From eq.~(\ref{eqnorm}) we deduce
that the modes $\Phi_q^-$ become normalizable for $\nu <1$. If $\nu$ is real,
this leads to two possible quantizations in the regime
\be -\frac{d^2} 4< m^2\ell^2 < -\frac{d^2} 4 +1\ee
related to each other by a Legendre transform~\cite{bibtach1,bibtach2}.

For $m^2\ell^2 < -d^2/4$, $\nu$ becomes imaginary and the unitarity bound
on the corresponding conformal field theory is violated. However, both modes
$\Phi_q^\pm$~(\ref{eqphi}) are normalizable under the inner product~({\em cf.}~(\ref{eqip}))
\be\label{eqipn} (\Psi_1^\pm,\Psi_2^\pm) = i\pi\int_{\Sigma_0} \frac{d^{d-1}x\,dz}{z^{d-1}} (\Psi_1^{\mp\star} \partial_0
\Psi_2^\pm - \partial_0 \Psi_1^{\mp\star} \Psi_2^\pm) \ee
The two modes are related to each other by complex
conjugation. Set
\be\label{eqnumu} \nu = i\mu\quad,\quad \mu = \sqrt{-m^2\ell^2 - \frac{d^2} 4} >0\ee
Another set of modes of interest are the Euclidean modes
\be\label{eqesa} \Psi_k^E = e^{ik_\mu x^\mu} \Phi_q^E (z)\;, \quad \Phi_q^E (z) = z^{d/2}
H_\nu^{(1)} (qz)\ee
which are linear combinations of the $\Phi_q^\pm$ modes.
The propagator for the modes $\Phi_q^+$~({\em cf.}~(\ref{eqgpsi}))
\be\label{eqgpsi2} G^+ (z,x^\mu;z',{x'}^\mu) = \int \frac{d^dk}{(2\pi)^d}\,
\Psi_k^{-\star} (z,x^\mu) \Psi_k^+ (z', {x'}^\mu)\ee
is given by the continuation of~(\ref{eqG}) to imaginary $\nu$.
The propagator for the $\Phi_q^-$ modes is then obtained by complex conjugation.
The Euclidean propagator, which corresponds
to the Euclidean modes~(\ref{eqesa}),
can be calculated after a Wick rotation $x^0\to -ix^0$,
which takes us to EAdS space. We obtain
\be\label{eqgH} G^E (z,x^\mu;z',{x'}^\mu) = \int \frac{d^dk}{(2\pi)^d}\,
e^{ik_\mu(x-x')^\mu}\; z^{d/2} K_\nu (q_Ez)\, {z'}^{d/2}
K_\nu (q_Ez')\ee
where $q_E^2 = k_0^2 + k_1^2 + \dots + k_{d-1}^2$.
After some algebra,
this can be brought into the form
\be\label{eqgH1} G^E (z,x^\mu;z',{x'}^\mu) = \frac{\Gamma(h_+)\Gamma(h_-)}{(2\sqrt\pi)^{d+1}\Gamma (\frac{d+1} 2)}\;
F(h_+,h_-; {\textstyle\frac{d+1} 2}; 1-P/4)\ee
where $P$ is the invariant distance given by~(\ref{eqP}) after a Wick rotation
on the time variable $x^0$.
The singularity is obtained by letting $P\to 0$, as before,
\be\label{eqsing1} G^E(z,x^\mu; z', {x'}^\mu) \sim \frac{\Gamma((d-1)/2)}
{4\pi^{(d+1)/2}} \; P_\epsilon^{-(d-1)/2}\ee
in agreement with our earlier result for modes with $m^2>0$~(\ref{eqsing}),
confirming the correct normalization of the Euclidean wavefunctions~(\ref{eqesa}).
It is also instructive to express
the Green function~(\ref{eqgpsi2}) corresponding to the choice of modes $\Phi_q^+$
(\ref{eqphi}) in terms of the Euclidean propagator. To this end, express $\Phi_q^+$ (in EAdS) in terms of the Euclidean modes,
\footnote{making use of the Bessel function identity
$I_\nu (x)= -\frac{i}{\pi} e^{-i\pi\nu} (K_\nu (x) - e^{i\pi\nu}\, K_\nu (-x))$.}
\be\label{eq33} \Phi_q^+ (z) = \mathcal{N} \left(\Phi_q^E (z) - e^{i\pi h_+}\; \Phi_q^E (-z)\right)\ee
where the normalization constant is
\be\label{eq37} \mathcal{N} = \frac{1}{\sqrt{\pi}} \; e^{\pi\mu /2}\ee
Then the Green function~(\ref{eqgpsi2})
may be straightforwardly expressed in terms of the Euclidean Green function~(\ref{eqgH1}).
Suppressing the (common) $(x^\mu, {x'}^\mu)$ dependence, we have
\be\label{eqGE} G^+ (z;z') = \mathcal{N}^2 \left( G^E (z;z')+ e^{2i\pi\nu} G^E (-z;-z')
- e^{i\pi h_+} G^E (z;-z')- e^{-i\pi h_-} G^E (-z;z')\right)\ee
It is straightforward to deduce the expression~(\ref{eqG}) for $G^+(z,z')$
from~(\ref{eqGE}) and (\ref{eqgH1}). Notice that all Green functions
in~(\ref{eqGE}) share the same $i\epsilon$ prescription, unlike the dS case~\cite{bibbms}.

Next, we discuss a dual approach by interchanging the roles of coordinates
and momenta. Thus, instead of imposing the gauge-fixing condition~(\ref{eqgf3})
on the coordinates $X^A$, we shall instead adopt the gauge
\be\label{eqc2a}
P_AP^A = -P_0^2 + P_1^2  + \dots + P_d^2 - P_{d+1}^2 = L^2
\ee
Eqs.~(\ref{eqcon4}) and (\ref{eqc2a}) imply
\be X_A X^A = -M^2\ee
where $\mathcal{C}_2 = -M^2L^2$ ({\em cf.}~eq.~(\ref{eqcon5})).
Note that this is identical to the condition~(\ref{eqgf3}) we imposed earlier
if $M^2 > 0$.

The Dirac brackets in this gauge are
\be\label{eqdir4} \{ X^A\;,\; P_B\}_D = \delta_B^A + \frac{1}{L^2} P^AP_B\ee
and this algebra may be realized by expressing the momenta in terms of
coordinates similar to the Poincar\'e coordinates~(\ref{eqpc}).
Then the coordinates $X^A$ will be given in terms of derivatives (conjugate
momenta) with respect to the Poincar\'e coordinates.

If the Casimir~(\ref{eqcas}) is positive ($\mathcal{C}_2 > 0$), then this
dual description leads to the same Scr\"odinger (wave) equation~(\ref{eqwe})
as before.
This is because $L^2 < 0$ in eq.~(\ref{eqc2a}), which makes it identical to
its dual counterpart~(\ref{eqgf3}) with the choice $\ell^2 = -L^2$.

In the case of a negative Casimir,  we have $L^2 > 0$ in eq.~(\ref{eqc2a}).
Upon analytic continuation of $P_{d+1} \to -iP_{d+1}$, eq.~(\ref{eqc2a}) becomes the definition of dS$_{d+1}$ space, albeit in momentum space,
\be\label{eqds3} -P_0^2 + P_1^2  + \dots + P_d^2 + P_{d+1}^2 = L^2\ee
We may express $P_A$ in terms of coordinates $(\widetilde x^\mu, \widetilde z)$
parametrizing the dS hyperboloid
as~({\em cf.}~eq.~(\ref{eqpc}))
\bea\label{eqpc1}
P_0 & = & - \frac {\widetilde z} 2 + \frac{\widetilde x^\mu \widetilde x_\mu + L^2}{2\widetilde z} \nonumber \\
P^i & = & \frac L {\widetilde z}\, \widetilde x^i\quad (i=1,\dots,d-1), \quad P_{d+1} = \frac L {\widetilde z}\, \widetilde x_0\nonumber \\
P^d & = & - \frac {\widetilde z} 2 + \frac{\widetilde x^\mu \widetilde x_\mu - L^2}{2\widetilde z}
\eea
where $\widetilde x^\mu \widetilde x_\mu = (\widetilde x^0)^2 +(\widetilde x^1)^2 + \dots + (\widetilde x^{d-1})^2$, which cover half of dS. The metric on dS reads
\be\label{eqmetr1}
ds_\mathrm{dS}^2 = L^2\; \frac{-d\widetilde z^2 +d\widetilde x^\mu d\widetilde x_\mu}{\widetilde z^2}
\ee
It can also be expressed in the more commonly used form
\be\label{eqmetr2}
ds_\mathrm{dS}^2 = L^2\; (-dt^2 +e^{-2t} d\widetilde x^\mu d\widetilde x_\mu)
\ee
by changing coordinates $\widetilde z = e^t$. Then the boundary $\widetilde z\to 0$ may be thought
of as the infinite past $t\to -\infty$. 
The invariant distance between
points $P^A$ and ${P'}^A$ ($A=0,1,\dots,d+1$) on the dS hyperboloid is
\be\label{eqP2} \widetilde P (P,P') = \frac{1}{L^2}\; (P'-P)^2 =
\frac{- (\widetilde z-\widetilde z')^2+(\widetilde x-\widetilde x')^\mu (\widetilde x-\widetilde x')_\mu}{\widetilde z\widetilde z'}\ee
Notice that the dS metric~(\ref{eqmetr1}) differs from the metric on EAdS (eq.~(\ref{eqmetr}) with Euclidean signature
for $dx^\mu dx_\mu$) in that $\widetilde z$ in dS is a timelike coordinate. This does not affect
the boundary behavior and the structure of the Green functions. It should also
be emphasized that there is no direct connection between the EAdS coordinates
$(z,x^\mu)$ and their dual counterparts $(\widetilde z, \widetilde x^\mu)$, even though there is a formal connection through double analytic continuation~\cite{bibadsds,bibadsds2,bibadsds3,bibbms},
because the latter parametrize the mass-shell hyperboloid~(\ref{eqc2a}) in the
embedding, whereas the former parametrize (E)AdS space.

The Casimir turns into the same wave equation as in EAdS~(\ref{eqwe}), since $\mathcal{C}_2 = -M^2L^2 =m^2\ell^2$ and $\partial_\mu\partial^\mu$ is the laplacian in ${\mathbf R}^d$.
The solutions are given by~(\ref{eqpw}) in terms of the Bessel functions~(\ref{eqphi}), where $q^2 = k_0^2+k_1^2+\dots +k_{d-1}^2$. Let us concentrate on the case of imaginary $\nu$ (i.e., $ML > d/2$; {\em cf.}~eq.~(\ref{eqnumu})).
The inner product in the space of solutions is similarly defined by~(\ref{eqip}), except that the spacelike slice $\Sigma_0$ should be defined as
$\widetilde z=$~const. The coordinates on the slice are $\widetilde x^\mu$. For the eigenfunctions~(\ref{eqpw}), the inner product reads
\be\label{eqip2}
(\Psi_1,\Psi_2) = i \int \frac{d^d\widetilde x}{\widetilde z^{d-1}}\; e^{i(k_1-k_2)_\mu \widetilde x^\mu}
(\Phi_1^\star(\widetilde z) \Phi_2'(\widetilde z)
- {\Phi_1^\star}' (\widetilde z) \Phi_2 (\widetilde z))\ee
The apparent $\widetilde z$-dependence disappears if we apply the wave equation~(\ref{eqwav}). The integral over $\widetilde x^\mu$ leads to a $\delta$-function, demonstrating the orthogonality of the wavefunctions.  For an othronormal set, choose
\be\label{eqos} \Phi_q(\widetilde z) \equiv \Phi_q^+ (\widetilde z) = C^+\, \widetilde z^{d/2} J_\nu (q\widetilde z)\ee
Using the Wronskian $J_\nu (x) J_{-\nu}' (x) - J_\nu' (x) J_{-\nu} (x) =
\frac{2i\sinh (\pi\mu)}{\pi x}$ (recall $\nu = i\mu$), we obtain
\be (\Psi_1,\Psi_2) = (2\pi)^d \delta^d (k_1-k_2) \; |C^+|^2 \; \frac{2\sinh (\pi\mu)}{\pi}\ee
showing that the set~(\ref{eqos}) is orthonormal if
\be C^+ = \sqrt\frac{\nu}{2} \; \Gamma (\nu)\ee
For the choice
\be\label{eqos2} \Phi_q(\widetilde z) \equiv \Phi_q^E (\widetilde z) = C^E\, \widetilde z^{d/2} H_\nu^{(1)} (q\widetilde z)\ee
using the Wronskian $H_\nu^{(1)} (x) {H_\nu^{(2)}}' (x) - {H_\nu^{(1)}}' (x) H_\nu^{(2)} (x) =
\frac{-4i}{\pi x}$, we similarly find
\be C^E = \frac{\sqrt\pi}{2}\; e^{-\pi\mu/2}\ee
The two orthonormal sets are related to each other by
\be\label{eq2pimu} \Phi_q^+ (\widetilde z) = \widetilde{\mathcal{N}}\; \left( \Phi_q^E (\widetilde z) -
e^{i\pi h_+} \Phi_q^E (-\widetilde z) \right)
\quad,\quad \widetilde{\mathcal{N}} = \frac{C^+}{2C^E} = \sqrt\frac{\nu}{2\pi}
\Gamma (\nu) e^{-i\pi\nu/2}\ee
where we used $J_\nu (x) = \half (H_\nu^{(1)}(x) -e^{i\pi\nu} H_\nu^{(1)} (-x))$.
This is in agreement with its dual counterpart~(\ref{eq33}) up to an overall
constant factor.
Notice also that~(\ref{eq2pimu}) differs from ref.~\cite{bibbms} by a phase factor
(note $|\mathcal{N}| = (1-e^{-2\pi\mu})^{-1/2}$). 
The Euclidean Green function is given by the same expression~(\ref{eqgH1}) as
before with $P$ replaced by $\widetilde P$ (\ref{eqP2}).
The Green function corresponding to the $\Phi_q^+$ modes may then be expressed
in terms of the Euclidean propagator. We obtain
\be\label{eqGE2} G^+ (z;z') = |\mathcal{N}|^2 \left( G^E (z;z')+ e^{2i\pi\nu} G^E (-z;-z')
- e^{i\pi h_+} G^E (z;-z')- e^{-i\pi h_-} G^E (-z;z')\right)\ee
which is of the same structure as the dual relation~(\ref{eqGE}). The two
normalization constants, $\widetilde{\mathcal{N}}$~(\ref{eq2pimu}) and
$\mathcal{N}$~(\ref{eq37}) differ and this is essential for the correct behavior
of the respective Green functions at the singularity. Unlike~(\ref{eqGE}),
the various Green functions entering~(\ref{eqGE2}) do not share the
same $i\epsilon$ prescription~\cite{bibbms}. This is because $\widetilde z$
is a timelike coordinate and the transformation $\widetilde z \to -\widetilde z$ is time reversal. This precludes a simple relationship between $z$ and
$\widetilde z$ by analytic continuation. As we have argued above, there exists
a duality transformation relating $(z,x^\mu)$ and $(\widetilde z, \widetilde x^\mu)$ by an interchange of coordinates and momenta in the $(d+2)$-dimensional
space in which (EA)dS space is embedded. The explicit form of this duality
transformation for $(z,x^\mu) \to (\widetilde z, \widetilde x^\mu)$ is rather
involved and uninspiring.

In conclusion, we have established a duality between tachyonic modes in
EAdS space and massive scalars in dS space in $d+1$ dimensions by interchanging the roles of
coordinates and momenta in the $(d+2)$-dimensional flat space in which (EA)dS space is embedded.
For massive modes in (E)AdS, this procedure leads to a self-duality. This
duality explains why in dS space one obtains
Green functions that are similar to their EAdS counterparts but for {\em tachyonic} modes, even though the two inner products are different, due to
the different roles of the timelike direction. It would be interesting to
extend the results to other modes and include spin. This is currently under
investigation.


\newpage

\begin{thebibliography}{99}
\bibitem{strominger}
A.~Strominger, hep-th/0106113.
\bibitem{bibds1}
M.~Li, {\tt hep-th/0106184}.
\bibitem{bibds2}
S.~Nojiri and S.~D.~Odintsov, Phys.~Lett.~{\bf B519} (2001) 145, {\tt hep-th/0106191}.
\bibitem{bibds3}
D. Klemm, Nucl.~Phys.~{\bf B625} (2002) 295, {\tt hep-th/0106247}.
\bibitem{bibds4}
S.~Nojiri and S.~D.~Odintsov, Phys.~Lett.~{\bf B519} (2001) 145, {\tt hep-th/0106191}.
\bibitem{bibds5}
S.~Nojiri and S.~D.~Odintsov, JHEP {\bf 0112} (2001) 033, {\tt hep-th/0107134}.
\bibitem{bibds6}
A.~Tolley and N.~Turok, {\tt hep-th/0108119}.
\bibitem{bibds7}
T.~Shiromizu, S.~Ida and T.~Torii, JHEP {\bf 0111} (2001) 010, {\tt hep-th/0109057}.
\bibitem{bibds8}
C. M. Hull, JHEP {\bf 0111} (2001) 012, {\tt hep-th/0109213}.
\bibitem{bibds9}
V. Balasubramanian, J. de Boer and D. Minic, {\tt hep-th/0110108}.
\bibitem{bibds10}
R.~Cai, Y.~Myung and Y.~Zhang, {\tt hep-th/0110234}.
\bibitem{bibds11}
U.~H.~Danielsson, {\tt hep-th/0110265}.
\bibitem{bibds12}
S.~Ogushi,Mod.~Phys.~Lett.~{\bf A17} (2002) 51, {\tt hep-th/0111008}.
\bibitem{bibds13}
A.~Petkou and G.~Siopsis, {\tt hep-th/0111085}.
\bibitem{bibds14}
A.~M.~Ghezelbash and R.~B.~Mann, JHEP {\bf 0201} (2002) 005,
{\tt hep-th/0111217}.
\bibitem{bibds15}
A.~J.~M.~Medved, Class.~Quant.~Grav.~{\bf 19} (2002) 919, {\tt hep-th/0111238}.
\bibitem{bibds16}
A.~Padilla, Phys.~Lett.~{\bf B528} (2002) 274, {\tt hep-th/0111247}.
\bibitem{bibds17}
A.~J.~M.~Medved, {\tt hep-th/0112009}.
\bibitem{bibds18}
M.~Cvetic, S.~Nojiri and S.~D.~Odintsov, {\tt hep-th/0112045}.
\bibitem{bibds19}
B.~McInnes, {\tt hep-th/0112066}.
\bibitem{bibds20}
Y.~S.~Myung, {\tt hep-th/0112140}.
\bibitem{bibds21}
S.~Nojiri and S.~D.~Odintsov, {\tt hep-th/0112152}.
\bibitem{bibds22}
A.~J.~M.~Medved, {\tt hep-th/0112226}.
\bibitem{bibds23}
A.~M.~Ghezelbash, D.~Ida, R.~B.~Mann, T.~Shiromizu, {\tt hep-th/0201004}.
\bibitem{bibds24}
E.~Halyo, {\tt hep-th/0201174}.
\bibitem{bibds25}
Y.~S.~Myung, {\tt hep-th/0201176}.
\bibitem{bibds26}
S.~Nojiri and S.~D.~Odintsov, {\tt hep-th/0201268}.
\bibitem{bibds27}
D.~Youm, {\tt hep-th/0201268}.
\bibitem{bibds28}
S.~R.~Das, {\tt hep-th/0202008}.
\bibitem{bibds29}
M.~Brigante, S.~Cacciatori, D.~Klemm, D.~Zanon, {\tt hep-th/020273}.
\bibitem{bibds30}
S.~Ness and G.~Siopsis, {\tt hep-th/0202096}.
\bibitem{bibads1}
J.~Maldacena,
Adv.~Theor.~Math.~Phys.~{\bf 2} (1998) 231, {\tt hep-th/9711200}.
\bibitem{bibads2}
G.~G.~Gubser, I.~R.~Klebanov and A.~M.~Polyakov,
Phys.~Lett.~{\bf B428} (1998) 105, {\tt hep-th/9802109}.
\bibitem{bibads3}
E.~Witten,
Adv.~Theor.~Math.~Phys.~{\bf 2} (1998) 253, {\tt hep-th/9802150}.
\bibitem{bibadsds}
C.~M.~Hull, JHEP {\bf 9807} (1998) 021, {\tt hep-th/9806146}.
\bibitem{bibadsds2}
V.~Balasubramanian, P.~Horava and D.~Minic, JHEP {\bf 0105} (2001) 043,
{\tt hep-th/0103171}.
\bibitem{bibadsds3}
B.~McInnes, {\tt hep-th/0110062}.
\bibitem{bibbms}
R.~Bousso, A.~Maloney and A.~Strominger, {\tt hep-th/0112218}.
\bibitem{bibtach1}
I.~R.~Klebanov and E.~Witten, Nucl.~Phys.~{\bf B556} (1999) 89, {\tt hep-th/9905104}.
\bibitem{bibtach2}
E.~Witten, {\tt hep-th/0112258}.




\end{thebibliography}
\end{document}